\documentclass[sigconf]{acmart}

\usepackage{amsmath}
\usepackage[usestackEOL]{stackengine}

\usepackage{xcolor}
\usepackage[inline,nomargin,index]{fixme}

\fxsetup{theme=color,mode=multiuser,inlineface=\itshape,envface=\itshape,status=draft}
\FXRegisterAuthor{mm}{amm}{\colorbox{gray!10!white}{\color{black}Mikaela}}
\FXRegisterAuthor{kl}{akl}{\colorbox{gray!10!white}{\color{black}Kristian}}
\FXRegisterAuthor{ah}{aah}{\colorbox{gray!10!white}{\color{black}Aaron}}
\FXRegisterAuthor{aem}{aem}{\colorbox{gray!10!white}{\color{black}Erica}}
\fxusetargetlayout{color}

\AtBeginDocument{%
  \providecommand\BibTeX{{%
    \normalfont B\kern-0.5em{\scshape i\kern-0.25em b}\kern-0.8em\TeX}}}

\copyrightyear{2022}
\acmYear{2022}
\setcopyright{rightsretained}
\acmConference[FAccT '22]{2022 ACM Conference on Fairness, Accountability, and Transparency}{June 21--24, 2022}{Seoul, Republic of Korea}
\acmBooktitle{2022 ACM Conference on Fairness, Accountability, and Transparency (FAccT '22), June 21--24, 2022, Seoul, Republic of Korea}\acmDOI{10.1145/3531146.3533104}
\acmISBN{978-1-4503-9352-2/22/06}

\begin{document}

\title[Flipping the Script on Criminal Justice Risk Assessment]{Flipping the Script on Criminal Justice Risk Assessment: An actuarial model for assessing the risk the federal sentencing system poses to defendants}

\author{Mikaela Meyer}
\email{mikaela@stat.cmu.edu}
\orcid{0000-0003-3718-3405}
\affiliation{%
  \institution{Department of Statistics \& Data Science and Heinz College, Carnegie Mellon University}
  \streetaddress{5000 Forbes Ave.}
  \city{Pittsburgh}
  \state{Pennsylvania}
  \country{USA}
  \postcode{15213}
}

\author{Aaron Horowitz}
\orcid{0000-0001-7931-8756}
\affiliation{%
  \institution{American Civil Liberties Union}
  \city{New York City}
  \state{New York}
  \country{USA}
}

\author{Erica Marshall}
\orcid{} 
\affiliation{
\institution{Idaho Justice Project}
\city{Boise}
\state{Idaho}
\country{USA}
}

\author{Kristian Lum}
\orcid{0000-0003-2637-5612}
\affiliation{
\institution{Department of Computer and Information Science, University of Pennsylvania}
\city{Philadelphia}
\state{Pennsylvania}
\country{USA}}


\begin{abstract}
  In the criminal justice system, algorithmic risk assessment instruments are used to predict the risk a defendant poses to society; examples include the risk of recidivating or the risk of failing to appear at future court dates. However, defendants are also at risk of harm from the criminal justice system. To date, there exists no risk assessment instrument that considers the risk the system poses to the individual. We develop a risk assessment instrument that ``flips the script.'' Using data about U.S. federal sentencing decisions, we build a risk assessment instrument that predicts the likelihood an individual will receive an especially lengthy sentence given factors that should be legally irrelevant to the sentencing decision. To do this, we develop a two-stage modeling approach. Our first-stage model is used to determine which sentences were ``especially lengthy.'' We then use a second-stage model to predict the defendant’s risk of receiving a sentence that is flagged as especially lengthy given factors that should be legally irrelevant. The factors that should be legally irrelevant include, for example, race, court location, and other socio-demographic information about the defendant. Our instrument achieves comparable predictive accuracy to risk assessment instruments used in pretrial and parole contexts. We discuss the limitations of our modeling approach and use the opportunity to highlight how traditional risk assessment instruments in various criminal justice settings also suffer from many of the same limitations and embedded value systems of their creators. 
\end{abstract}




\maketitle

\section{Introduction}
Algorithmic risk assessment instruments are used within the criminal justice system in settings such as pretrial release, bail determinations, sentencing, and parole supervision \cite{garrett2019assessing,Viljoen2019-zj,administrative2011overview, cohen2018federal, Bao2021-uc}. The objective of a risk assessment instrument is to estimate the likelihood that an individual will face an adverse outcome. In the pretrial context, risk assessment instruments generally predict to what degree an individual is at risk of committing a new crime or failing to appear at future court dates if they were to be released on some conditions as opposed to being detained until their court date \cite{mdemichele2018public}. In this context, these risk predictions inform whether someone should be released pre-trial and, if so, under what conditions. In the context of sentencing, recidivism predictions can influence the sentence a person receives. Once a person has been incarcerated, similar recidivism prediction tools are often used to influence whether they will be released on parole or to justify levels of parole supervision \cite{Brennan2009-ph}.

In most instances in criminal justice settings of which we are aware, the intended purpose of the risk assessment instrument is to estimate the risk posed by the defendant. For example, the estimated likelihood of re-arrest is often communicated as the risk the accused person poses to public safety.  However, there is substantial evidence that the system also poses risks to the defendant. These risks include being denied pretrial release, receiving a sentence disproportionately lengthy for the given conviction, and being wrongfully convicted \cite{Digard2019-eo, schmitt2016overview, garrett2020wrongful}. A tool that reverses the goal estimates the risks posed by the system to the individual. Such a tool could be used to prepare defense attorneys to best defend their clients, warn individuals who are likely to be treated unjustly by the system, notify  judges about the risk to ensure cases receive appropriate review, and inform post-sentencing motions for compassionate release. However, to our knowledge, the literature about tools that flip the script in this way is limited. Recent work by Barabas et al. \cite{barabas2020studying} proposed a judicial risk assessment that predicts whether a judge will set a bail amount that is unaffordable for an individual. Smith et al. \cite{smith2021racial} studied differences in racial disparities in sentencing across U.S. federal judges, and Helsby et al. \cite{helsby2018early} developed a data-driven approach to predict which police officers will have an adverse interaction with a member of the public. To our knowledge, a tool that makes predictions for individual defendants in the criminal legal system does not yet exist.

In this paper we develop a risk assessment instrument to fill this gap. Specifically, using a dataset consisting of U.S. District Court sentences from 2016 to 2017 \cite{ciocanel2020justfair}, we build a risk assessment instrument to predict the likelihood that a defendant in U.S federal court will receive an ``especially lengthy'' sentence. To do this, we use a two-stage modeling approach. In the first-stage model, we use a Bayesian non-parametric model to estimate the conditional distribution of sentence length given legally relevant factors for each case. We label a sentence as having been especially lengthy if it is greater than the estimated 0.9 quantile of its conditional distribution. Informally, we label sentences as especially lengthy if the model estimates they would be in the top 10\% of all cases sharing the same legally relevant factors. These binary labels serve as the predictive target of the second-stage model. The second-stage model then predicts whether an individual is at risk of receiving an especially lengthy sentence based on legally irrelevant factors. These irrelevant factors include the defendant’s socio-demographics as well as the judge’s socio-demographics and details about which part of the country the defendant is being tried in. 

We begin with background information on sentencing in U.S. federal courts, including the process by which an individual is sentenced and the laws and policies governing sentencing decisions in Section \ref{sec:background}. In Section \ref{sec:data}, we describe the data used to build our model. Section \ref{sec:methods} describes our modeling approach. Section \ref{sec:results} gives model results, including validation of the models. Section \ref{sec:limitations} explains technical limitations, while Section \ref{sec:values} discusses the normative choices we made during model development, highlighting how these also apply to existing ``unflipped'' risk assessments.  We conclude in Section \ref{sec:conclusion}.  


\section{Background on Federal Sentencing}\label{sec:background}
In this section, we describe the federal court system, paying particular attention to sentencing.

\subsubsection{The Landscape} 

The federal criminal code is set forth in a number of statutes in federal code. Title 18, entitled Crimes and Criminal Procedure, criminalizes the broadest swath of conduct and also sets forth laws about the procedure of criminal cases in the United States \cite{title18_undated-wb}. Other conduct is criminalized in more topic-specific sections of the United States Code. 
The federal crimes are policed by federal law enforcement agencies, such as the Federal Bureau of Investigation, and ultimately prosecuted by the United States Attorney’s Office in one of the 94 federal court jurisdictions across the country. The United States federal court system is divided into 94 individual jurisdictions based on population, where each state has at least one federal court jurisdiction. For example, Pennsylvania contains three jurisdictions, including the Eastern, Western, and Middle Districts of Pennsylvania, while Idaho is comprised simply of the District of Idaho. Each of the 94 jurisdictions has its own United States Attorney, which represents the federal government’s chief prosecutor for that jurisdiction. Each year, federal district judges around the country sentence thousands of criminal defendants.  Yearly sentence totals have ranged from 67,742 individuals sentenced in 2016 \cite{schmitt2016overview} to 86,201 individuals sentenced in 2011 \cite{reedt2011overview}. In 2019, there were 90,473 new federal criminal cases filed \cite{caseload-stats-2019}.

\subsection{Steps in a Case}
A case begins when the U.S. Attorney files either a Criminal Complaint or an Indictment against the individual, at which point an arrest warrant will be issued (if the person is not already in custody). From here, the defendant will make an initial appearance, file any pre-trial motions, and may negotiate a plea deal with the prosecutor. In 2016, 97\% of cases were resolved by a plea bargain \cite{schmitt2016overview}. If no plea deal is reached, the case will proceed to trial where either a judge or jury will determine whether the defendant is guilty or innocent.  

After a determination of guilt—-through either a plea bargain or trial—-the defendant will begin the sentencing process. The sentencing process starts with a review of the Federal Sentencing Guidelines. The Federal Sentencing Guidelines are a large manual that includes a point system to be applied to each unique offense to reflect the severity of the crime and any mitigating circumstances.  This number is displayed as an offense level. For example, a person convicted of a drug crime is subject to the provisions of Chapter 2, Part D of the Sentencing Guidelines \cite{chapter2guidelines_2021-au}. This Chapter establishes a base offense level of anywhere between 6 and 43, depending on drug type, quantity, whether bodily injury resulted, and other factors. Once the base offense level is determined, the defendant receives additional points for any ``specific offense characteristic.'' Two points each are added for factors such as any threat of violence in commission of the crime, use of a firearm, use of mass marketing, maintaining a premises for the purposes of distribution, or committing the offense as part of a livelihood, to name just a few. Additional points are then added or subtracted pursuant to ``adjustments'' made to account for the victim of the crime, the defendant’s role in the offense (mitigating or aggravating), and other conduct. 

\begin{figure*}[h]
\includegraphics[scale = 0.8]{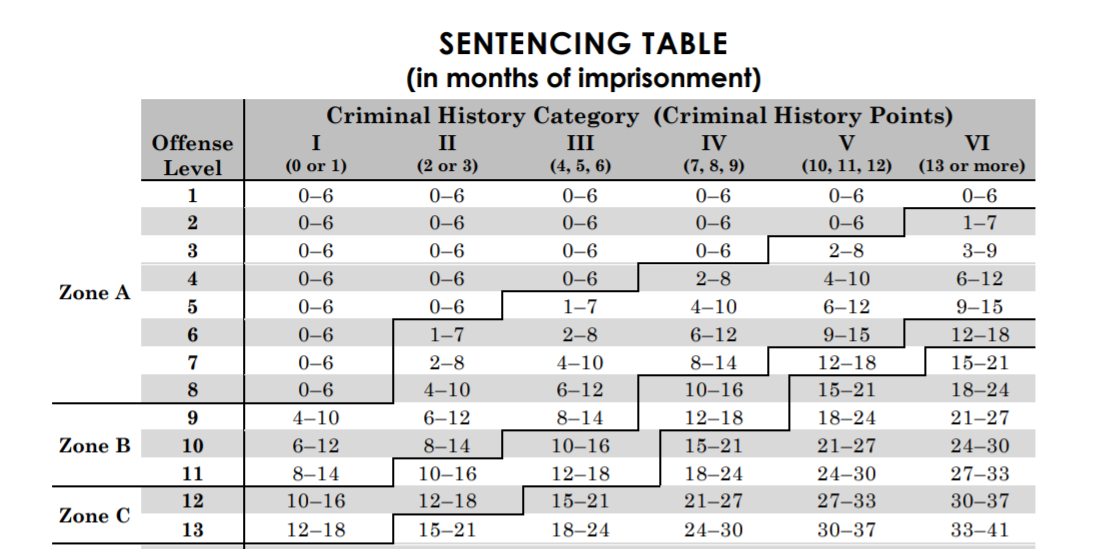}
\caption{\label{fig:guidelines} An excerpt of a chart in \cite{chapter5guidelines_2021} for determining the sentencing guideline range. Offense levels continue below up to 43, and sentences can be as long as ``life.''}
\end{figure*}

The Sentencing Guidelines then use a point system to count prior convictions of the offender. Points are also added if a defendant committed the offense while under any criminal justice sentence, including probation, parole, or supervised release. This number is displayed as the criminal history category.  The final offense level can be paired with the defendant's criminal history category to determine the range of months of imprisonment, known as the sentencing guideline range, as seen in Figure \ref{fig:guidelines}. 

In most cases, the United States Probation Office performs an evaluation of the case, calculates the sentencing guideline range, conducts an interview with the defendant, and sets forth its recommended sentence to the judge in a Presentence Investigation Report (``PSR''). Both parties file sentencing memoranda responding to this PSR and stating their position on an appropriate sentence length. 

The sentencing process culminates with a sentencing hearing before a judge, typically the same judge that has had the case from its inception. Because the judge has the discretion to levy sentences, in most jurisdictions, the judge, not the terms of any plea bargain or the recommendations in the PSR, will determine the sentence. At the conclusion of the sentencing hearing, the judge will make legal determinations about which Sentencing Guideline provisions govern, determine the applicable sentencing guideline range, discuss reasoning for any upward or downward departure or a variance from the sentencing guideline range, and order the prison sentence, in months, as well as any terms of probation. For more details about the federal sentencing process, see \cite{sentencing-basics}.
\subsection{Sentencing Disparities and Previous Literature}
For nearly twenty years, the Guidelines were mandatory, meaning that federal judges had no discretion to deviate from the sentencing range specified under the Guidelines \cite{yang2014have}.
However, in 2005, the Supreme Court’s decision in United States v. Booker rendered the Sentencing Guidelines advisory \cite{starr2013mandatory}. 
Since Booker, federal sentencing disparities have been the topic of extensive research and have been widely documented. Many published papers have concluded that two similarly-situated defendants (meaning two defendants with similar criminal histories and similar offense conduct) end up with different sentences due to factors that should be legally irrelevant, such as race, \cite{Yang2015-tf, Tuttle2019-oe,Smith2021-au}, gender \cite{starr2015estimating}, the jurisdiction the defendant is prosecuted in, the appointment date of the sentencing judge, and the appointing President of the sentencing judge \cite{yang2014have,Cohen2019-ro}, to name a few. In 2017, the United States Sentencing Commission concluded that similarly-situated Black men received, on average, a sentence that is 19.1\% longer than their white counterparts \cite{United_States_Sentencing_Commission2018-ej}.  

While the Sentencing Commission has proposed a return to mandatory Guidelines \cite{starr2013mandatory}, the Guidelines have been criticized by scholars and practitioners for their ``rigidity,'' and for shifting the power to prosecutors in their charge and plea bargaining decisions \cite{Yang2015-tf}. Other scholars have pointed out that it was not Booker and its progeny that created the disparities, but that different sentencing outcomes based on race are part of the criminal legal system from its inception, starting with policing decisions about who is arrested and charging decisions made by prosecutors through their use of prosecutorial discretion \cite{Starr2013-zi}.   

To date, sentencing disparities persist even though the law requires courts to consider the ``need to avoid unwarranted sentence disparities among defendants with similar records who have been found guilty of similar conduct'' \cite{no-sent-disp}. While there are many studies documenting the sentencing disparities after they occurred, to the best of our knowledge, there is no study or tool that takes a forward-looking approach to predict the risk that a defendant is likely to experience disparate treatment in their particular case. The risk assessment instrument we develop in this paper aims to predict disproportionately harsh sentences prior to sentencing with the goal of hopefully avoiding these disproportionate sentences and reducing the disparities in the system going forward.

\section{Data}\label{sec:data}

Our model is built from data on U.S. federal court sentences. We use the JUSTFAIR dataset, which was created by \cite{ciocanel2020justfair} and made publicly available. JUSTFAIR merges publicly available data on federal sentences from the U.S. Sentencing Commission with data from the Federal Judicial Center, the Public Access to Court Electronic Records (PACER) system, and Wikipedia. Thus, for each sentence in JUSTFAIR, there is information about the nature of the charge(s), the details of the sentence, and the demographics of the defendant and the judge. The dataset covers about 600,000 sentences from 2001-2018, though as explained in \cite{ciocanel2020justfair}, not every federal sentence given in that time period is included in JUSTFAIR because data is lost throughout the merging pipeline. For example, about 50,000 sentences are dropped from the dataset when trying to add judge initials. 

We restrict the JUSTFAIR dataset to only include the 57,080 sentences that were given in the years 2016 and 2017. We cleaned the data such that for any categorical variable that had NA values, we created an ``unknown'' category or assigned NA values to this already existing ``unknown'' category. For some variables that referred to the number of offense level enhancements that someone was given based on a charge for a particular category (e.g., points given based on whether or not the crime was a drug crime), we assumed that NA values could be converted to zero points from that category. 



We divide the covariates in the JUSTFAIR dataset into two categories: legally relevant ($X$) and legally irrelevant ($Z$) factors. The relevant factors include:

\begin{itemize}
    \item the criminal history of the defendant
    \item the offense for which the defendant is being charged (number of counts for conviction, number of unique statutes in a case, whether there is a charge for using or possessing a firearm in the course of drug trafficking or a violent crime, etc.) 
    \item the sentencing guideline range and the adjustments made to determine this range (the baseline offense level used as a starting point when doing a guideline computation, which year's Guideline Manual was used to make the guideline computation, how many points were subtracted because the defendant accepted responsibility, how many points were added because a victim of the offense was a law enforcement or corrections official, which Chapter Two Guideline was eventually applied, etc.)
    \item the statutory minimum and/or maximum sentence lengths determined by the nature of the charge (e.g., the minimum sentence someone could get because there is at least one count associated with a drug or gun statute)

\end{itemize}

The decision about which variables to include as legally relevant was made by the team after reviewing the codebook \cite{ussc-codebook} for factors that were specific to the offense conduct and criminal history, and therefore would be relevant to calculations under the U.S. Sentencing Guidelines. These factors could fairly or legally be considered by the judge in issuing the sentence. We included as many legally relevant factors as possible but decided to omit certain factors that would have significantly reduced the size of the dataset because they were missing values for a large proportion of sentences in the dataset. 

The legally irrelevant factors include:

 \begin{itemize}
    \item socio-demographic information about the defendant (race, ethnicity, age, gender, U.S. citizenship status, and highest level of education)
    \item socio-demographic and political information about the judge (race, gender, political party of appointing president, etc.)
    \item information about the location, timing, and status of the case (whether there was a plea agreement, the month of the sentence, the judicial circuit and district in which the case was handled, and whether certain administrative documents were obtained) 
\end{itemize}

These factors are deemed legally irrelevant because they are not pertinent to the charge for which the individual is being sentenced. Individuals who have different values for any of these factors but the same values for the relevant factors would ideally receive the same sentence.

For an outcome variable ($Y$) for our first-stage model, we use sentence length in months (SENTTOT0 in the JUSTFAIR dataset). To prevent the long upper tail of sentence lengths from influencing our first-stage model's results, we cap the outcome variable at 540 months, which is approximately equal to the 99.9th percentile of sentence lengths. Life sentences, which are automatically set to 470 months for this variable in the original dataset, are set equal to this capped value, too. After these data cleaning measures are taken, we end up with 56,430 sentences with all legally relevant factors present and 55,580 sentences with all legally irrelevant factors present. We train our models on 80\% of the data and reserve 20\% for validation. 

\section{Methods}\label{sec:methods}

We develop a two-stage process for building our risk assessment instrument. In summary, the first-stage model determines which sentences were ``especially lengthy'' relative to the legally relevant factors given in the sentencing guideline calculations. The output of the stage 1 model -- a binary indicator -- is the outcome variable predicted in the second stage, which is the final risk assessment instrument. In stage 2, similar to many RAIs, we use a generalized linear model. In this model, we include only legally irrelevant factors as covariates to estimate the likelihood that an individual will receive one of the especially lengthy sentences conditioned only on factors that should not be considered in sentencing. Below, we give the details of each of the models. 

\subsection{Stage 1 Model: Predict sentence length and flag ``especially lengthy'' sentences}

To build a model that predicts which individuals will receive especially lengthy sentences, we must first know which sentences are especially lengthy. That is, we need an outcome variable. In traditional risk assessments, the outcome variable is available by processing administrative records. For example, a common outcome variable is whether an individual is re-arrested during some defined time period. While extracting this information from the systems in which it is stored may not be trivial, it does not require inference or modeling. In our case, because data that mark sentences as having been especially lengthy do not exist, we begin the process of building our risk assessment model by inferring which sentences were especially lengthy. We note that the first-stage model is unnecessary in traditional risk assessments because the outcome variable in traditional risk assessments can be calculated without modeling. 

To infer which sentences were especially lengthy, we need to know what the distribution of sentence lengths is for any given set of legally relevant factors. There are typically few sentences that share precisely the same set of legally relevant factors, so simply using the empirical distribution of sentence lengths for each set of cases sharing the same legally relevant factors is not possible. Thus, we require good model-based estimates of the conditional {\it distribution} of $Y$ (the sentence length, in months)  given $X$ (the legally relevant factors). Traditional regression approaches assume that the distribution of the errors is constant and independent of $X$. In our case, we expect that the variance of the errors is non-constant and likely varies with $X$. For example, we expect that sentence lengths for cases where the Guidelines indicate a minimal sentence are less variable in terms of the absolute number of months than sentences for those cases where the Guidelines indicate a long sentence. We also expect that legally relevant factors influence judicial decision-making in complicated ways that are not well approximated by a simple linear function. 

Heteroscedastic Bayesian autoregressive trees (HBART) \cite{pratola2017heteroscedastic} offers a flexible, non-parametric approach designed to model both nonlinearity in the mean function and heteroscedasticity in errors.  HBART models both $E[Y|x]$ and $\text{Var}[Y|x]$ using a ``sum-of-trees'' approach for the conditional expectation and a ``product-of-trees'' approach for the conditional variance. Specifically, HBART assumes that a process $Y(x)$ can be represented as:
\begin{equation}
    Y(x) = f(x) + s(x) \xi
    \label{eq:hbart}
\end{equation}
where $f(x) = E[Y|x]$, $s^2(x) = \text{Var}[Y|x]$, and $ \xi \sim N(0, 1)$. In words, HBART fits a nonlinear regression function, $f(x)$ to estimate the conditional expected sentence length given covariates. Simultaneously, it estimates the conditional variance of sentence length, $s^2(x)$, given those same covariates. Fitting is done via MCMC where at each iteration, we obtain draws from the posterior distribution of $f(x)$ and $s(x)$ for each individual. From these samples, we extract $\bar{f}(x)$ and $\bar{s}(x)$ -- the posterior means across all MCMC samples of $f(x)$ and $s(x)$, respectively.  

We define a sentence as having been an ``especially lengthy'' sentence if the sentence that was handed down was greater than the upper bound of the individual’s $(1-\alpha)\times 100\%$ one-sided predictive interval. For example, if $\alpha = 0.1$, we flag a sentence as especially lengthy if the model indicates that $90\%$ of all sentences for cases with identical legally relevant factors would receive lower sentences. In mathematical terms, we define 

\begin{equation}
Y^* = \mbox{1}[Y(x) > \bar{f}(x) + \Phi^{-1}(\alpha)\bar{s}(x)]
\label{eq:flag}
\end{equation}

\noindent where $\Phi^{-1}$ is the inverse cumulative distribution function of the standard normal distribution. This indicator variable is then used as the outcome variable in our second-stage model. 

For our purposes, we let $\alpha = 0.1$, and in doing so note that the choice to define especially lengthy sentences as the top 10\% is arbitrary. One could also choose the top 25\% or 5\%, for example. In practice, those different values of $\alpha$ did not result in substantial differences to the predictive accuracy of our second-stage model. 

\subsection{Stage 2 Model: Predict whether an individual is at risk of receiving an especially lengthy sentence}
For our second-stage model, we estimate the likelihood of an especially lengthy sentence (as defined in the first-stage model) given legally irrelevant covariates. 
Specifically, we estimate a logistic regression model:

\begin{equation}
    \text{Pr}(Y^* = 1) = logit^{-1}(Z\beta).
    \label{eq:logit}
\end{equation}

Under this model, each of the legally irrelevant covariates ($Z$) receives a weight given by the corresponding element of $\beta$. These weights are multiplied by each of the factors, summed together, and transformed to the probability scale via the inverse logit function. We estimate the $\beta$ parameters using LASSO as implemented in \texttt{glmnet} \cite{glmnet-citation}. This estimation approach uses regularization.  We select the regularization parameter using 10-fold cross-validation, selecting the degree of regularization that maximizes the out-of-sample area under the ROC curve (AUC). {\bf This model defines our final risk assessment model.} It predicts the likelihood that an individual will receive an especially lengthy sentence based only on factors that should be legally irrelevant.

\section{Results}\label{sec:results}
We review the results of the first- and second-stage models in turn.

\subsection{Stage 1 Model}
\begin{figure*}
    \centering
    \includegraphics[scale = .8]{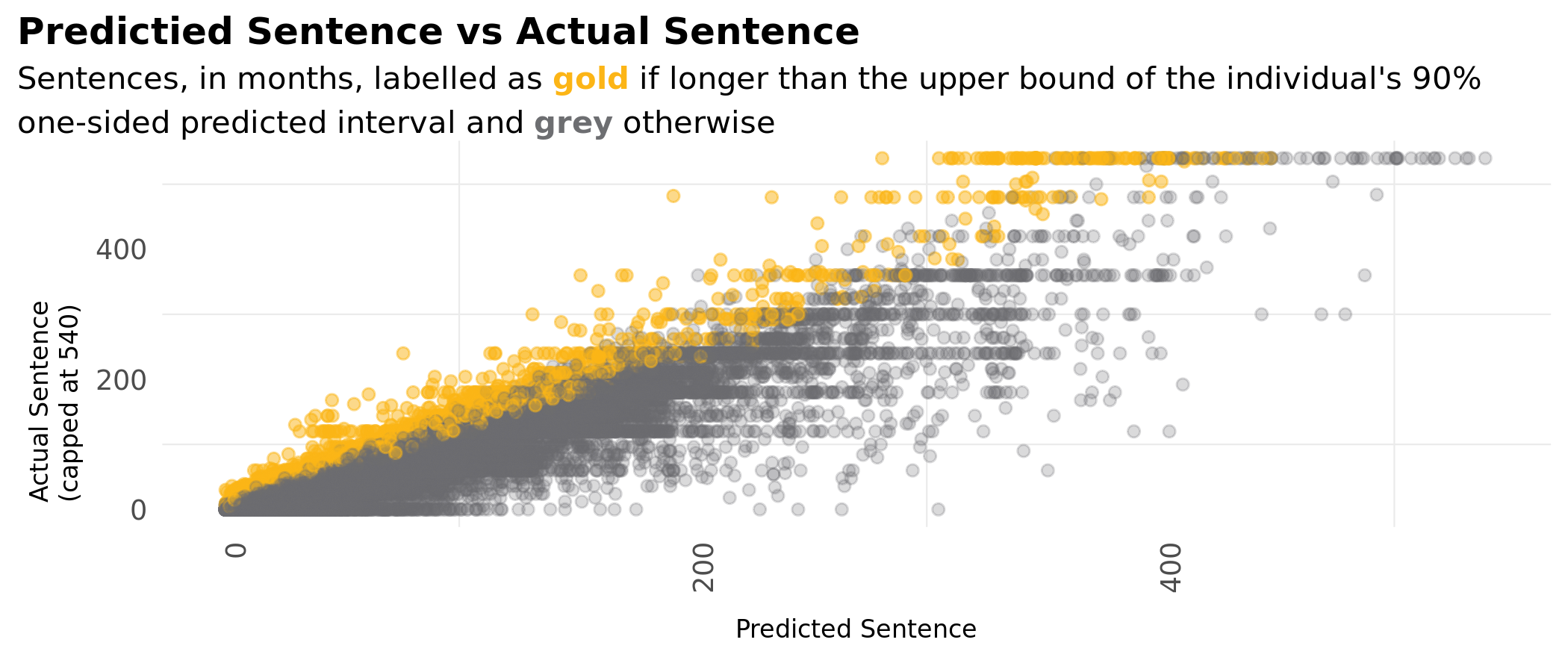}
    \caption{Predicted sentences from our Equation \ref{eq:hbart} model, color coded by Equation \ref{eq:flag}, vs. actual sentences. Sentences are capped at 540 months in the dataset.}
    \label{fig:fig1}
\end{figure*}

We fit our first-stage model using the \texttt{rbart} package in \texttt{R} \cite{rbart-citation}. We run the MCMC algorithm for 10,100 iterations, discarding the first 100 iterations as burn-in. Trace plots and convergence diagnostics for the MCMC are given in Appendix \ref{sec:stage1app} (Figures \ref{fig:mcmc_mean} and \ref{fig:mcmc_sd}). The adjusted R-squared for the training and test sets is 0.9 and 0.89, respectively, which shows that the legally relevant factors are, unsurprisingly, able to predict much of the variation in sentence length, and the model is not over-fitting. Figure \ref{fig:fig1} shows the actual sentence length versus the predicted sentence length. Sentences that fall above the estimated 0.9 conditional quantile are shown in gold. These are the sentences we flag as especially lengthy. Sentences that were not flagged as especially lengthy are shown in grey.  

One observation that may stand out is that there are examples where there are two cases with very similar predicted sentences (two dots that appear near each other on the horizontal axis), but the longer sentence is not flagged as especially lengthy while the shorter one is (the grey dot is higher than the gold dot).  This is not an error in the model, but rather a consequence of the fact that our model flexibly accounts for heteroscedasticity. That is, these are cases where there are two sets of legally relevant factors for which the model estimates similar {\it average}  sentences for both but different variances. 

This figure also highlights the importance of modeling the error variance directly. We see that -- as we expected -- for sentences that we predict to fall on the shorter end of the spectrum, the sentences are much more tightly clustered. For sentences that we predict to be near zero, there are no cases where our prediction is off by, say, 100 months. However, for cases where the model predicts a very long sentence, predictions that differ from the truth by 100 months do occur. Had we implemented a model that assumed a constant error variance as in traditional regression approaches, we would have flagged all sentences as especially lengthy that upwardly deviated from their predicted value by the same amount. This would have resulted in flagging none of the sentences with short predicted sentences and substantially more than the desired 10\% of sentences on the other end of the spectrum. In practice, this would lead us to fail to flag a (hypothetical) instance in which one case received a five year sentence and all other cases with the same legally relevant factors received a one month sentence. Because we want to flag sentences as especially lengthy {\it relative to similar cases}, this would be undesirable. 

\begin{figure*}
    \centering
    \includegraphics[scale = 0.9]{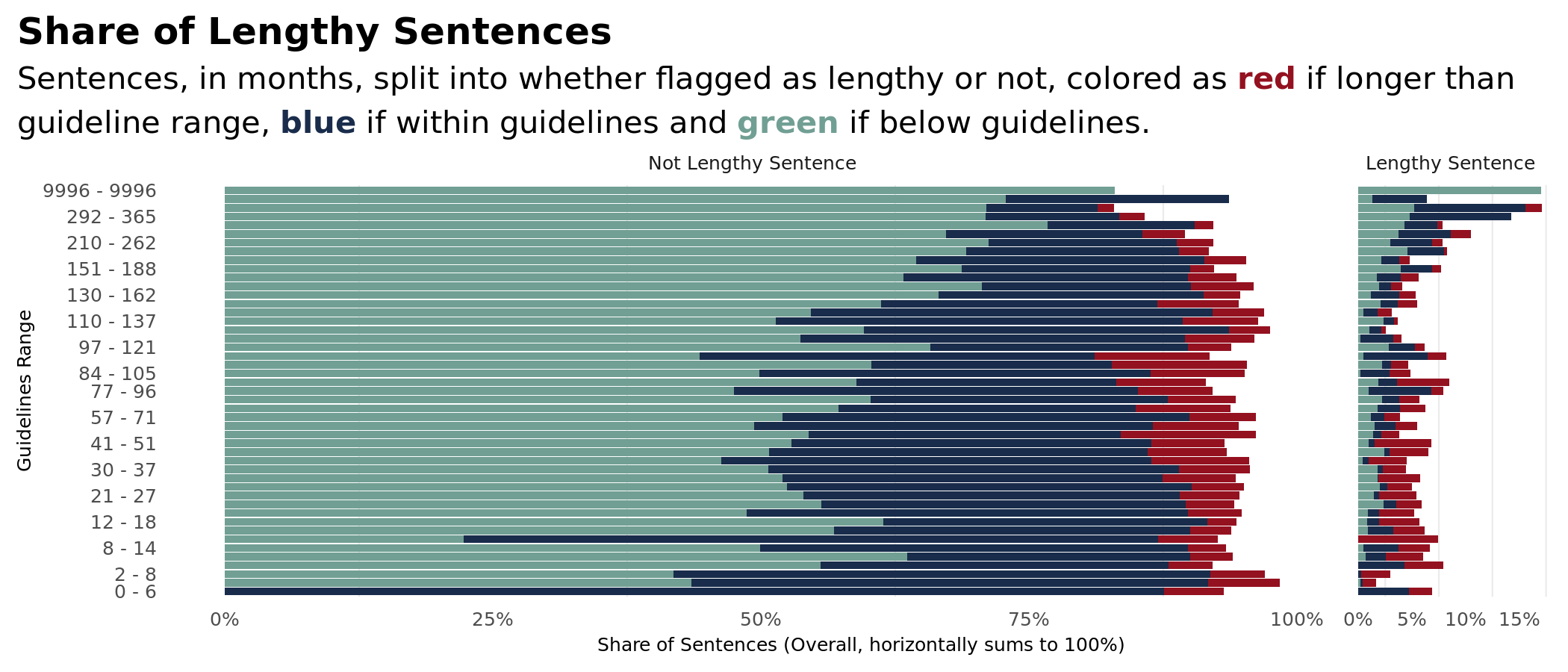}
    \caption{Left: The proportion of sentences within each Title 18 Sentencing Guideline Range which our Equation \ref{eq:flag} model does \emph{not} classify as especially lengthy, color coded by whether below, within, or above guidelines. Right: The proportion of sentences that are flagged by Equation \ref{eq:flag} as being especially lengthy binned by guideline range, with same color coding as left figure.}
    \label{fig:fig2}
\end{figure*}

Figure \ref{fig:fig2} shows the proportion of sentences within/outside of sentencing guideline ranges binned by guideline range. The left graph shows these proportions for the sentences that were not flagged as especially lengthy, while the proportions for sentences flagged as especially lengthy are displayed in the graph on the right. We see several encouraging results.  We flag sentences above, within, and below guideline ranges, indicating our model is picking up relevant factors beyond guideline ranges and can be useful even in cases where guideline ranges are met/exceeded -- and especially for lower sentencing guideline ranges (e.g., 0-6 months) that are rarely exceeded but can still be especially lengthy. It is clear the judges use substantial discretion and do not adhere strictly to the guideline ranges; thus using the guideline ranges alone may be insufficient to derive ``especially lengthy'' sentences. We also see that -- although we do flag less than the desired 10\% of sentences for most bins, indicating that the model slightly over-estimates the conditional variance -- we flag a similar proportion of sentences across the spectrum of sentencing guideline ranges, with the exception of longer sentencing guideline ranges which are less frequent. The fact that the sentences on the longest end of the ranges are flagged at a higher rate is likely a consequence of the capping at 540 months to account for life sentences and death sentences, as these are all calculated to have received the same sentence in months. This is reassuring that we are not excluding cases like the above example from being flagged. Overall, we flag about 6\% of sentences in both the training and the test sets as especially lengthy. 

\subsection{Stage 2 Model}

Our second-stage model, which is the risk assessment instrument itself, is fit using the \texttt{glmnet} package in \texttt{R} \cite{glmnet-citation}. The variables and coefficient values are summarized in Table \ref{tab:stage2coef} in Appendix \ref{sec:stage2app}. Figure \ref{fig:roc-bins} examines the performance of our second-stage model in two ways. On the right, we plot the ROC curve for the test set. As reported in Table \ref{tab:model2-aucs}, our out-of-sample AUC is 0.64. Based on standards previously established for criminal justice risk assessments by \cite{Desmarais2013-df}, our model's AUC falls in the range of 0.64-0.7, which is considered ``good.'' AUCs from 0.55-0.63 are considered ``fair,'' while AUCs greater than 0.71 are labeled ``excellent'' \cite{Desmarais2013-df}. For context, \cite{Desmarais2013-df} also found that COMPAS, a frequently-used recidivism prediction risk assessment instrument, has a median AUC across various studies that is considered ``good'' by their standards. Also, the PSA, a popular pretrial risk assessment instrument, was found to have an AUC that is also considered ``good'' by the same standards using pretrial data from Kentucky \cite{mdemichele2018public}. Thus, our instrument performs comparably to other instruments that are currently being used in various criminal justice decision-making processes. Recall we chose to flag actual sentences as especially lengthy if they were longer than 90\% of all sentences for cases with identical legally relevant factors. Even if we defined sentences as especially lengthy if they were longer than 75\% of sentences for cases with the same legally relevant factors, our instrument would still have been considered to be borderline between ``fair'' and ``good,'' and this was the worst case scenario for our instrument's performance. 

We also visualize the proportion of sentences that are flagged as especially lengthy based on an individual's predicted risk category in the left graph of Figure \ref{fig:roc-bins}. The five predicted risk categories were defined such that an equal number of cases fell in each. Thus, people in the high risk category have a predicted probability of receiving an especially lengthy sentence that is in the top 20\% of all predicted probabilities. Though we do not include confidence intervals for these proportions in this figure, we note that these intervals are tiny (the largest standard errors for the proportions is 0.0001) due to the large number of observations (about 2,220) that fall in each bin, and the risk categories' intervals do not overlap with each other. Based on this figure, an individual in the high risk category is empirically about four times as likely to receive an especially lengthy sentence than someone who is in the low risk category. Further, the base rate of especially lengthy sentences in the test set is 6\% across all categories, so the proportion of especially lengthy sentences in the high risk category is about twice as high as the base rate across all categories. This difference between groups is similar to differences observed in other tools that bin scores to create risk groups. To illustrate, \cite{mdemichele2018public} found that for the outcome of failing to appear (FTA) for the Public Safety Assessment when tested on Kentucky's pretrial data, individuals in the two top risk categories of six categories had empirical FTA rates that were twice the base FTA rate. For the outcome of new violent criminal activity (NVCA), \cite{mdemichele2018public} observed that individuals flagged by the Public Safety Assessment for having a higher likelihood of committing a violent crime during pretrial release were three times as likely to actually commit new criminal activity than those who were not flagged. We also note a monotonic increase in the empirical rate of lengthy sentences with the risk categories in our validation set. This pattern is often sought and provides evidence of predictive validity of an assessment instrument, as in  studies of the Public Safety Assessment \cite{mdemichele2018public}, the federal Post-Conviction Risk Assessment (PCRA) instrument \cite{cohen2018federal}, and COMPAS \cite{flores2016false}.

\begin{figure*}
    \centering
    \includegraphics[scale = 0.9]{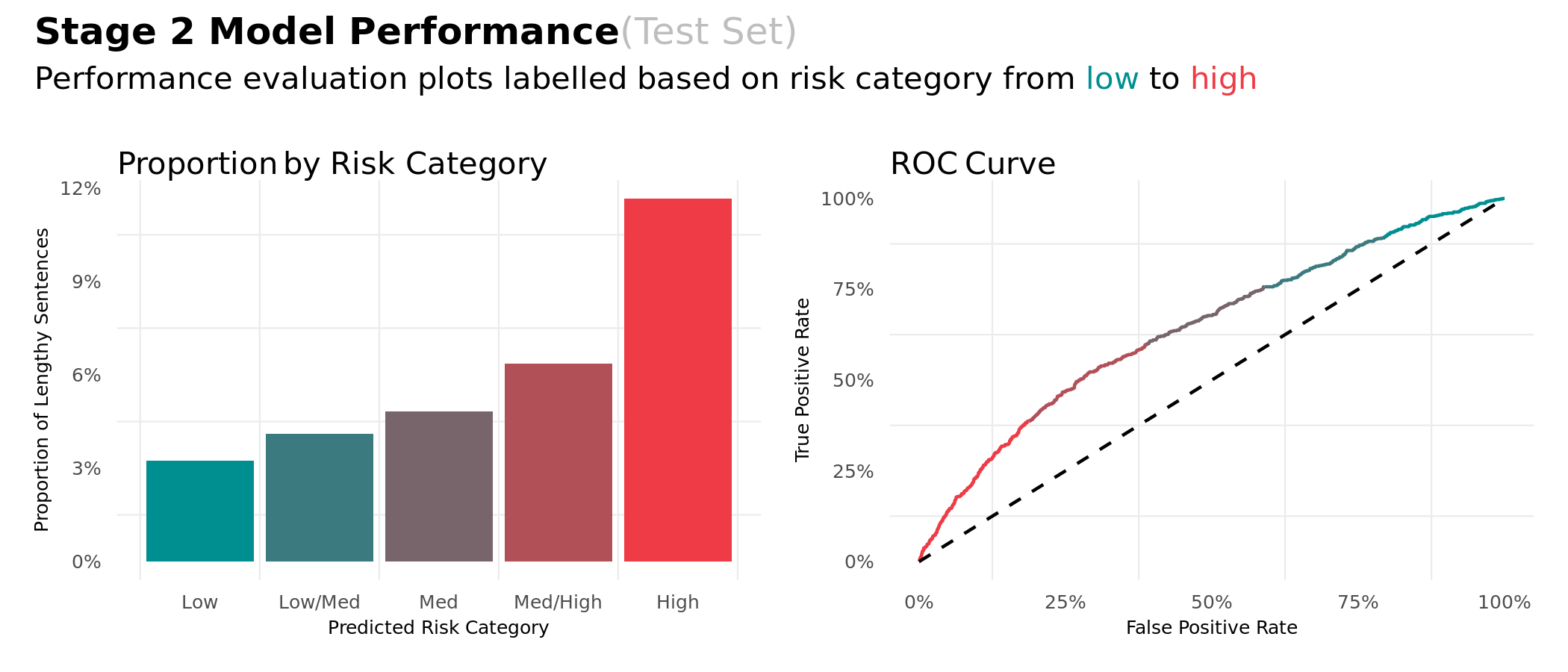}
    \caption{Left: Empirical performance of Equation \ref{eq:logit} on the test set of binned predicted risks from ``Low'' to ``High'' in equivalent sized bins, as is common among risk assessments. Color-coded from lowest to highest risk. Right: ROC curve for the test set, with regions color-coded based on the predicted risk category from the left figure}
    \label{fig:roc-bins}
\end{figure*}

\begin{table}
\begin{tabular}{|c|c|c|}
\hline
 & \textbf{{Training Set}} & \textbf{{Test Set}} \\ 
 & \textbf{{AUC}} & \textbf{{AUC}} \\ 
\hline
\textbf{10\% Flagged} & 0.66                                                                & 0.64                                                             \\ \hline
\textbf{15\% Flagged} & 0.65                                                                & 0.64                                                              \\ \hline
\textbf{20\% Flagged} & 0.63                                                                & 0.63                                                             \\ \hline
\textbf{25\% Flagged} & 0.64                                                                & 0.63                                                             \\ \hline
\end{tabular}
\caption{In-sample and out-of-sample AUCs based on the top percentage of sentences that are flagged as especially lengthy (the $\alpha$ value in Equation \ref{eq:flag})}
\label{tab:model2-aucs}
\end{table}

\section{Technical limitations}\label{sec:limitations}


Our model is not without shortcomings and limitations. 
Many of these shortcomings stem from the data on which the model is built. For example, upstream policy decisions or limitations in the processes and management of the criminal justice system impact which data is available or not \cite{Pfaff2017-at}. In our case, while presence or absence of pleas are available, the details of the plea bargaining process itself are not considered in our models. As discussed earlier, estimates for the number of federal guilty pleas resolved via plea bargaining range from 90-95 percent \cite{bjs_05, flanagan_m90, pastore_m03} and there is substantial evidence of plea bargaining processes impacting case outcomes, known as the ``trial penalty'' \cite{Ulmer2006-zi,Ulmer2010-ph,Devers2011-gt}. Evidence also suggests there are meaningful plea bargaining impacts based on an individual's race \cite{Tuttle2019-oe}, jail status at time of bargain \cite{Gupta2016-hd,Leslie2017-wm}, prior arrests \cite{Luka_undated-ew}, and more. Our limited knowledge of the details and data from plea bargaining are caused in part by the closed-door nature of plea bargaining activities and the poor tracking of charges as they evolve from initial charges to charges ultimately applied at sentencing \cite{schneider2019bargaining, turner2020transparency}. 

Data used in criminal justice risk assessments have similar problems with poor data tracking and quality. For instance, many pretrial risk assessments rely heavily on past failures to appear in court to predict future failures to appear. But there are many different reasons for failing to appear and different types of proceedings people failed to appear to that are not considered. Much like the details of plea bargaining, this information would add important context to the data. In a validation and critique of the CPAT/CPAT-R in Colorado, it was noted that a single failure to appear in the prior year weighed heavily on the estimated risk for a given defendant, almost guaranteeing pretrial detention recommendations \cite{Wallace2020-au}. During validation, the auditors could evaluate the consequences of the FTAs as the court viewed them (e.g., warrants, sanctions, etc). 28\% of FTAs were no or low-consequence, even by the court's own punitive standards, but because the risk assessment lacks these details, the risk assessment places anyone with any FTA in the last year in a high risk category \cite{Terranova2020-pq}.

The data we use for our model are also not the totality of information available to human decision-makers at sentencing decisions. While much of this information is contained in PACER, this system currently charges per-case access. Some of the data from PACER is included in JUSTFAIR, but it is not included in its entirety \cite{ciocanel2020justfair, Martin2008-or}. Additionally, there is information a judge might have access to in making a sentencing decision -- perhaps information they have learned during the trial -- that is not systematized and recorded anywhere and thus not accounted for by our model. Traditional risk assessments are also limited in the same way. For example, in the pretrial context, judges can speak with the defendant, their attorney, or the assistant district attorney (ADA) to obtain more information relevant to their decision that is not available to any pretrial risk assessment instrument in the future.

Our dataset also suffers from sampling bias; people within this dataset are not necessarily representative of all sentenced people because of a mix of our own filtering procedures and choices made prior to our use of the dataset by the JUSTFAIR study authors \cite{ciocanel2020justfair}. For example, the dataset we rely on does not contain cases for several jurisdictions throughout the country even though it is a federal dataset. By comparison to traditional criminal justice datasets, however, we believe sampling bias in the JUSTFAIR dataset is considerably less problematic \cite{Bushway2007-uk, Berk1983-zi}. In typical pretrial settings, the datasets only contain people who have been released, creating substantial sampling bias. 
Similarly, in parole/probation applications, if people are rearrested for technical violations, they are often left-censored (reincarcerated), and we are not able to realize the counterfactual of being rearrested (or not) because of supervision activity \cite{Minor2003-cp, Grattet2016-tw}. 

Another major issue with our approach concerns construct validity. In particular, there is no undisputed ``ground truth'' for what constitutes an especially lengthy sentence \cite{Jacobs2021-rk, Jacobs2021-og}. Our definition of especially lengthy sentences is based on current judicial behaviors in the U.S. federal system. This definition embeds the assumption that the status quo in this system is more or less reasonable -- that all but the most extreme sentences are not too lengthy. Certainly, sentencing practices and expectations vary widely between the U.S. and other similar countries \cite{Subramanian_undated-od, Penal_Reform_International2018-om}, and substantial evidence suggests negligible (and possibly negative) returns to longer sentences for rehabilitation purposes and limited returns for deterrence purposes as well \cite{Loeffler2022-ru,Mastrobuoni2016-os,Lee2005-qe}. A reasonable person could argue that the definition of especially lengthy sentences should not be pegged to baselines determined by the norms of the current system but rather to the standards of other countries or normative beliefs about the value of lengthy sentences. Alternatively, people could consider our approach too harsh on the sentencing system, noting that we are missing critical context that judges can perceive such as the full rap sheet of an offender, or that sentence length on its own may not be the only indicator of outcomes given that other punishments can result from judicial choices.  

All that said, similar issues with measurement of the outcome variable and construct validity exist in traditional criminal justice risk assessment. For example, in pretrial decision making, \cite{1987united} dictates that public safety and willful flight are the only two justifications for pretrial detention, and detention should be a carefully limited exception \cite{makingsenseofriskassessment}.  However, neither the concept of public safety nor willful flight are especially measurable constructs \cite{Slobogin2003-ou,Mayson_dangerous_defendants,Gouldin2018-oc}. A broad definition of public safety may include concerns about the public health effects of short stays in jail while awaiting trial \cite{Lofgren2020-bk, Reinhart2021-tt, Kajeepeta2021-cr}, the heightened risk of incarceration itself (illness, injury, death, etc.) for people \cite{Victor2021-ib,Trotter2018-qh,mentalillnesstatesurv_2014-oe,Venters2019-su, Wildeman2018-jl}, the likely financial and emotional harm to families and loved ones \cite{Comfort2016-zd, The_Annie_E_Casey_Foundation_focus_children, Trusts2010-ss}, the risk of time spent in jail inducing future criminal activity \cite{Gupta2016-hd,Leslie2017-wm,Heaton2016-ex}, and more follow-on effects \cite{Turney2019-wk} that make our society less safe. In practice, most modern pretrial tools focus on the likelihood that a person will be rearrested as a proxy for the concept of public safety even though the vast majority of rearrest activities are not for violence \cite{mdemichele2018public}, and even violence-focused rearrests are not reliable proxies for conviction of offenses \cite{Fogliato2021-xy}. Similarly, most models operationalize willful flight as predicting whether the defendant will fail to appear for any future court appointments without distinguishing the reason why they failed to appear, which could be any of a variety of non-flight reasons \cite{Gouldin2018-oc}. In both cases, there is a mismatch between what is used as the outcome variable and the concept that is intended to be predicted. 

A related measurement issue arises for the variables we use as covariates. For example, by conditioning on the ``legally relevant factors'' in our first-stage model, there is an implicit assumption that these factors as recorded are accurate and fair representations of each person's criminal conduct, etc. However, there is evidence that the factors may themselves exhibit differential bias along important social or demographic lines. For example, \cite{Tuttle2019-oe} presents evidence of prosecutorial gaming. Under newly established sentencing guidelines, there was a substantial shift in the proportion of drug-related cases for which the weight of the illicit drugs in question fell just above the threshold for harsher treatment. This shift was particularly pronounced for Black and Hispanic defendants. Traditional risk assessment instruments use input factors that also are questionable with respect to the extent to which they offer an unbiased measurement of a person's conduct. Questions remain as to the validity of arrest as an unbiased measure of offense \cite{Fogliato2021-xy}.

Also, because we train our models on sentences from 2016 and 2017, their predictive power and goodness-of-fit are reliant on some degree of stationarity in the legal system. For example, amendments are sometimes made to the federal Sentencing Guidelines, which could change the sentencing range for a particular crime (the minimum or maximum guideline suggestion) or the number of levels that are added to or subtracted from a defendant's calculation \cite{ussg-amend-2016, ussg-amend-2018}. This limitation is not unique to our risk assessment; other risk assessment instruments also face this issue when changes are made that lead to potential distribution shift or covariate shift \cite{Koepke2018-ak}. Risk assessment instruments like ours are trained on historical data and used to make predictions with future data, so if policy or other changes are made that affect underlying distributions in the model, the instruments will likely give less accurate predictions.

Our two-stage modeling approach presents some limitations that have no analogue in traditional risk assessments because traditional risk assessment models do not need a first-stage model. In using this approach, our second-stage model's performance and predictions rely heavily upon the performance of our first-stage model and choices made to train this model (e.g., which type of model to use, setting any necessary parameters, etc.). Though our first-stage model fits the data well, our second-stage model is limited by the degree to which our first-stage model captures or does not capture the appropriate sentence length for a case with certain legally relevant factors. This is because we do not have access to true values that our first-stage model is estimating, such as the true values of $f(x)$ or $s(x)$ in Equation \ref{eq:hbart}. Furthermore, our decision to dichotomize a continuous value -- the residuals of our first-stage model -- results in a loss of information. However, we chose to use a binary outcome variable in our second-stage model to closely mirror traditional risk assessment instruments, which tend to predict binary outcomes.

\section{Embedded values choices}\label{sec:values}

Modeling risk in a sociotechnical context requires a number of choices to be made by developers of these tools, which are often left unacknowledged and unexamined \cite{Lehr_undated-aq,Dobbe2019-ms}. 
To emulate and demonstrate the extensiveness of these choices, our modeling approach attempts to adhere to modeling evaluation standards and guidelines \cite{Hanson2010-bo, Rajlic2010-le,Skeem_undated-wd, Helmus2017-yd, Singh2013-kd, Singh2015-gy} in the comparable criminal justice risk assessments literature \cite{Desmarais2013-df, Singh2018-lr,Desmarais2021-bb, Desmarais2016-sp}. Our interdisciplinary team of researchers, advocates, and lawyers made similar choices that would be required when developing typical criminal justice risk assessments. The make up of our team is itself a limiting factor that matches risk assessment development: the choices made by our team for a proposed tool for use in the service of the public are highly undemocratic, centralizing of our own perspectives, and generating of policy absent effective process, if used without substantial additional public deliberation \cite{Okidegbe2020-tt,Mulligan2019-mg,Alexandrova_undated-mx}. 

Risk assessment designers have substantial discretion in problem formulation and modeling decisions that are highly subjective \cite{Passi2020-makingdswork,Muller2019-humanstudyds,Passi2019-problemform}. For example, choosing the cut point for defining especially lengthy sentences in our instrument is akin to choosing how to measure outcomes in other risk assessments. Also, similar to other types of risk assessment instruments, we made decisions about which variables to include in our model and how to pre-process them. These decisions are based on our own assumptions and understanding of what is legally or morally relevant or irrelevant, what is worth including, and technical details like how to treat missing values. 

Modeling choices also embed values. 
Purely model-based variable selection techniques embed the belief that a single accuracy metric (typically, some measure of predictive accuracy) justifies variable choices and is the sole objective to optimize against. On the other end of the spectrum, variable selection may also be done by model builders who manually select variables based on their preferences for which variables ought to be included or excluded from the model based on moral, ethical, or theoretical considerations. Perhaps more obviously, this process also embeds the creators' values. Like other risk assessment instruments,  our approach falls between these two extremes. For example, after initially training our second-stage model but before looking at coefficients associated with predictors, we realized we included a predictor that we {\it a priori} believed to be unrelated to unjustified disparities in sentence lengths. We checked our second-stage model's AUC both before and after removing this variable, and we found that the out-of-sample AUCs were approximately the same. Though both models have approximately the same AUC, we prefer the model without this predictor in it.

Finally, our accuracy metric, AUC, inherently assumes equal valuation of false positives and negatives, which may be inappropriate in this context \cite{Flach2011-au,Lobo2008-uv}. Equal valuation of false positives and negatives certainly may not be the case in other criminal justice risk assessments, where \cite{Stevenson2021-fr} demonstrated the steep costs of pretrial incarceration are not internalized by current systems. Yet, AUC is the standard; therefore equal valuation of false positives and negatives remains the status quo for criminal justice risk assessments \cite{Fazel_undated-la, Helmus2017-yd}. 



\section{Conclusion}\label{sec:conclusion}
We have created a risk assessment instrument that predicts a defendant's risk of receiving an especially lengthy sentence based on factors that should be legally irrelevant in determining the sentence length. Our instrument performs comparably to other risk assessment instruments used in the criminal justice setting, and the predictive accuracy it achieves is considered ``good'' by the standards of the field. 
Our first-stage and second-stage models, either independently or in combination, can be used by various actors to advocate for reducing over-sentencing in a federal context. The original intent -- to shift power from prosecutors and judges to public defenders and defendants -- could be achieved by providing risk scores for especially lengthy sentences and distributional information about sentencing decisions to these groups to allow them to make informed decisions about navigating sentencing proceedings and plea bargaining. Using our tool properly could create a positive feedback loop that reduces disparities in sentencing as individual sentences are evaluated and potentially adjusted.

We see two other potentially exciting applications. First, in 2018, Congress passed, and the President signed into law, the First Step Act.  This law enabled defendants, for the first time in history, to directly file motions with the court to seek a sentence reduction where ``extraordinary and compelling'' circumstances warrant a reduction \cite{first-step-act}. Since then, interpreting the statutory text of the First Step Act, federal district courts across the country have granted thousands of such motions where the personal history of the defendant, the underlying offense, the original sentence, the disparity created by any changes in the law, and the sentencing factors at section 3553 warrant such a reduction. With even just the first of our two models, defendants could point to how far their sentence deviates from what we would predict based on the characteristics of their case, including the ability to look at how their case may be resolved today versus when they were sentenced. With the second-stage model, petitioners could point to the legally irrelevant factors that may have influenced their sentencing.
 
Second, the presidential power of clemency is set forth in the U.S. Constitution in Article II, section 2. This power allows the President to grant pardons to individuals convicted of federal crimes. Using either model to support creating clemency lists could prove beneficial. Currently, clemency is an underutilized presidential power, mostly used only for specific cases, to provide relief to classes of people sentenced under now-defunct criminal laws and/or charges society deems the current system overly punitive for (e.g., non-violent drug offenses) \cite{Larkin2016-wo}. All of these use cases leave out most federally incarcerated people for even the slim possibility of mercy. As our models perform well across the distribution of sentence lengths, this approach overcomes those limitations and can provide suggestions for relief, even for people often left out of criminal justice reforms like those who have been sentenced for violent crimes or sex offenses \cite{Pfaff2017-at}.

Finally, in interrogating the limitations and values choices inherent to the construction of our own model, we have highlighted many parallel issues with traditional risk assessment instruments as \cite{barabas2020studying} also reflected upon. We expect that there will be objections to the use of our model on the basis of these technical limitations and the subjectivity of the choices that were made. To the extent that one may reasonably believe that these limitations make our model unsuitable for use to estimate the risk {\it to} defendants, we would hope that such objections would be equally applied to the question of the suitability of similar such models to estimate the risk posed {\it by} defendants.

\begin{acks}
This material is based upon work supported by the the National
Science Foundation Graduate Research Fellowship Program under
Grant No. DGE1745016.
\end{acks}

\bibliographystyle{ACM-Reference-Format}
\bibliography{sample-base}

\appendix


\section{Stage 1 Model}\label{sec:stage1app}

\begin{figure}[h]
\vspace*{-10pt}
    \centering
    \includegraphics[width=\linewidth]{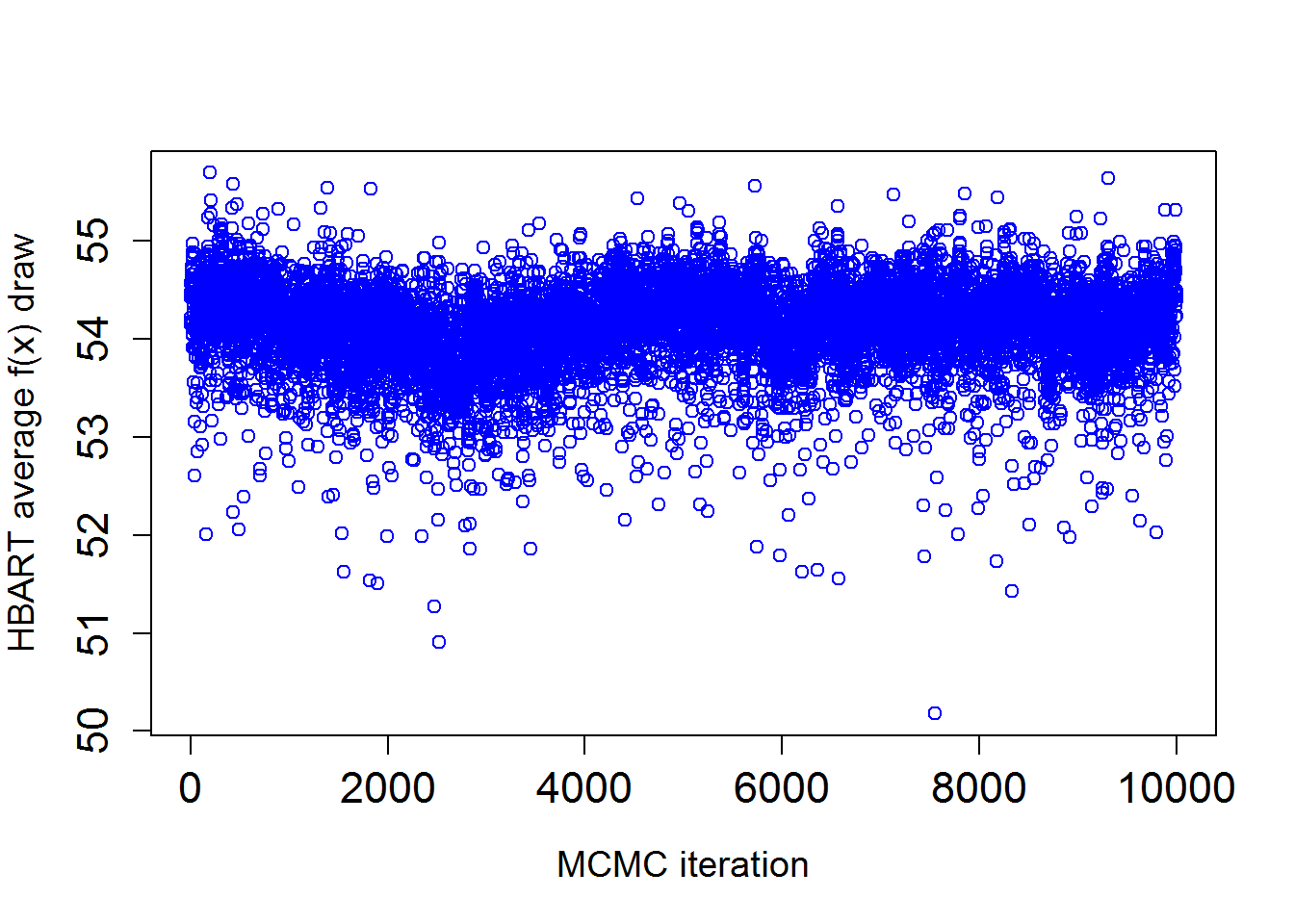}
    \caption{Average draw of mean function, f(x), across all individuals for each MCMC iteration}
    \label{fig:mcmc_mean}
\end{figure}

We used the Geweke diagnostic \cite{geweke1992evaluating} to test whether the Markov chain has converged to the stationary distribution. The z-score for this test for $\bar{f(x)}$ is $1.183 (\text{p-value} = 0.237)$, so there is evidence that the Markov chain converged to its stationary distribution.

\begin{figure}[h]
\vspace*{-10pt}
    \centering
    \includegraphics[width=\linewidth]{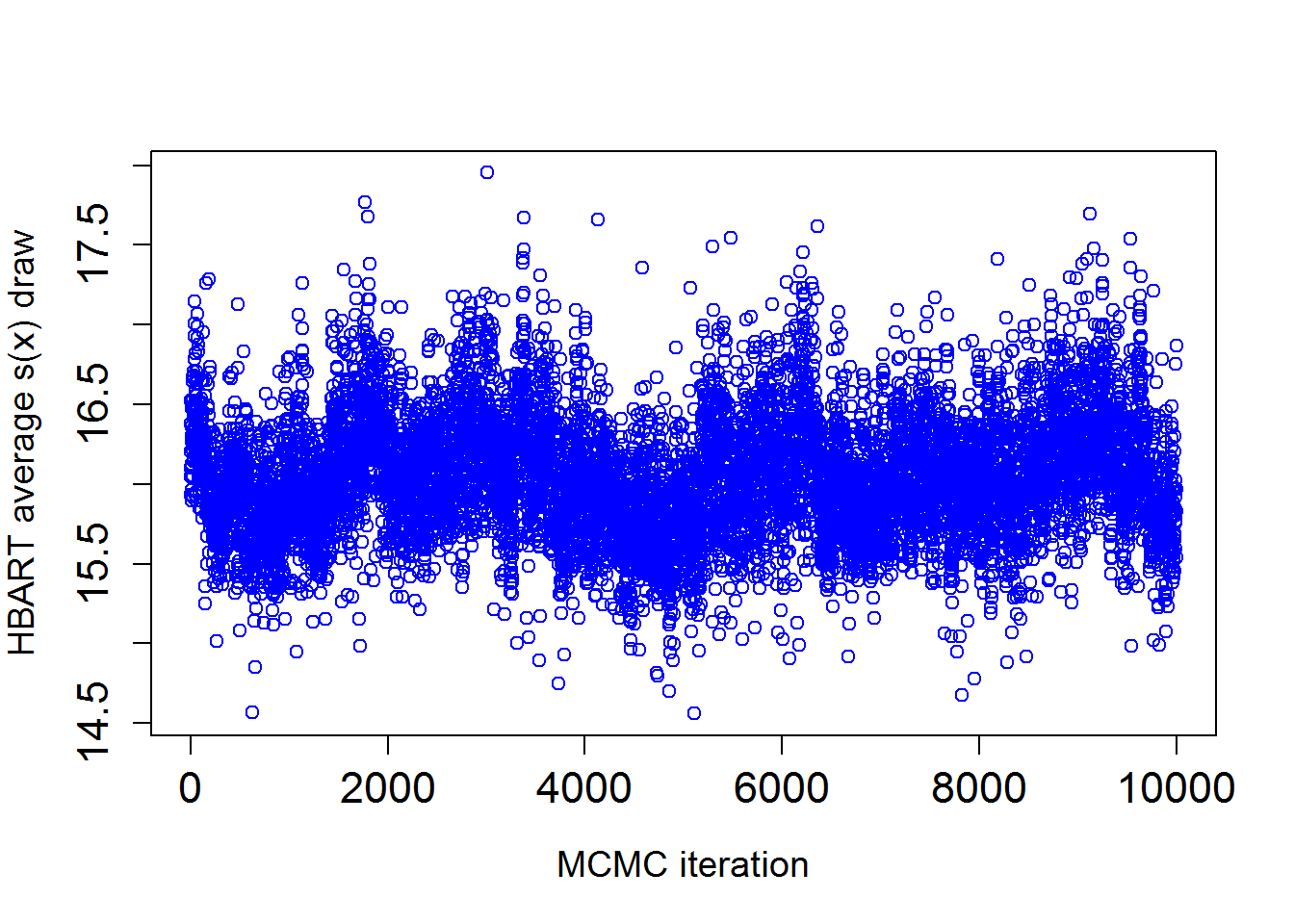}
    \caption{Average draw of standard deviation function, s(x), across all individuals for each MCMC iteration}
    \label{fig:mcmc_sd}
\end{figure}

We again used the Geweke diagnostic \cite{geweke1992evaluating} to test whether the Markov chain has converged to the stationary distribution. The z-score for this test for $\bar{s(x)}$ is $-0.845 (\text{p-value} = 0.4)$, so there is evidence that the Markov chain converged to its stationary distribution.

\section{Stage 2 Model} \label{sec:stage2app}

\onecolumn

\begin{table}
\begin{tabular}{|l|c|c|c|c|}
\hline
\textbf{Description} & \textbf{Variable Name} & \textbf{{Number of }} & \textbf{Min.} & \textbf{Max.} \\ 
 & & \textbf{{Non-Zero}} & \textbf{{Coefficient}} & \textbf{{Coefficient}} \\
 & & \textbf{{Coefficients}} & & \textbf{{(if not equal to Min.)}}\\ \hline
Defendant's race & MONRACE & 1 & 0.191 &  \\ \hline
{{Document status}} &  &  & &\\ {{indicator for Plea}} & DSPLEA & 0 & & \\ {{Agreement}} & & &  &  \\ \hline
Month of sentence & SENTMON & 0 &  &  \\ \hline
{{Disposition of}} & & & & \\ {{defendant's case}} & DISPOSIT & 1 & 0.253 &  \\ \hline
{{Offender's pre-sentence}} & & & & \\ {{detention status}} & PRESENT & 2 & -0.352 & -0.314 \\ \hline
{{Document status}} & & & & \\ {{indicator for Indictment}} & DSIND & 1 & 0.808\\ {{or Information}} & & & &  \\ \hline
{{Defendant's ethnic}} & & & & \\ {{origin}} & HISPORIG & 1 & 0.038 &  \\ \hline
{{Judicial circuit where}} & & & & \\ {{sentenced}} & MONCIRC & 4 & -0.04 & 0.123\\ \hline
{{Document status}} & & & & \\ {{indicator for Pre-}} & DSPSR & 1 & -0.56 &\\ {{Sentence Report}}& &  & &  \\ \hline
{{District where}} & & & &\\ {{sentenced}} & DISTRICT & 11 & -0.062 & 0.817 \\ \hline
{{Defendant's age when}} & & & &\\ {{sentenced}} & AGE & 0 &  &  \\ \hline
Defendant's gender & MONSEX & 1 & -0.004 &  \\ \hline
{{Defendant's citizenship}} & & & & \\ {{with respect to U.S.}} & NEWCIT & 0 &  &  \\ \hline
{{Defendant's highest}} & & & &\\ {{level of education}} & NEWEDUC & 1 & -0.003 &  \\ \hline
{{Party of President that}} & & & &\\ {{appointed judge}} & PartyofAppointingPresident1 & 1 & 0.005 &  \\ \hline
{{Indicator for whether}} & & & &\\ {{judge began serving}} & PreBooker & 0 & &\\ {{before Booker decision}} & & &  &  \\ \hline
Judge's race/ethnicity & RaceorEthnicity & 0 &  &  \\ \hline
Judge's gender & Gender & 0 &  &  \\ \hline
{{Interaction of}} & & & & \\ {{defendant's race and}} & MONRACE*RaceorEthnicity & 0 & & \\ {{judge's race/ethnicity}} & & &  &  \\ \hline
{{Interaction of}} & & & & \\ {{defendant's gender and}} & MONSEX*Gender & 0 & & \\ {{judge's gender}} & & &  &  \\ \hline
{{Interaction of circuit}} & & & & \\ {{and party of appointing}} & MONCIRC* & 5 & -0.195 & 0.173\\ {{President}} & PartyofAppointingPresident &  &  & \\ \hline
Year sentenced & SENTYR & 0 &  &  \\ \hline
\end{tabular}
\caption{Table of coefficients for stage 2 model on the logit scale. Many variables are treated as factors, so for each level of these factor variables, there is a possible coefficient in the model. Further descriptions of variables can be found in \cite{ussc-codebook} and \cite{ciocanel2020justfairdict} with the exception of ``PreBooker'', which is a binary variable that we created to denote whether a judge began serving before (1) or after (0) the Booker ruling.}
\label{tab:stage2coef}
\end{table}

\end{document}